\documentstyle[aas2pp4]{article}

\begin{document}

\title{Off-center deflagrations in Chandrasekhar mass SN Ia models}
\author{J.C. Niemeyer, W. Hillebrandt, and S.E. Woosley\altaffilmark{1}}
 
\affil{Max Planck Institut f\"ur Astrophysik, Karl Schwarzschild
Str.1, 85740 Garching, Germany}

\altaffiltext{1}{permanent address: Board of Studies in Astronomy and
Astrophysics, University of California, Santa Cruz, CA 95064} 

\authoraddr{Max Planck Institut f\"ur Astrophysik, Karl Schwarzschild
Str.1, 85740 Garching, Germany}

\newcommand{\ea}{{\it et al.~}}
\newcommand{\be}{\begin{equation}}
\newcommand{\ee}{\end{equation}}
\newcommand{\mc}{M$_{\rm chan}$~}
\newcommand{\cms}{cm s$^{-1}$}

\begin{abstract}

A series of two dimensional numerical simulations of explosive
nuclear burning is presented for white dwarfs near the Chandraskhar mass.
We assume that the burning begins as a slow deflagration front at or near the
center of the star, and continues until the density in the burning regions
has declined to about 10$^7$ g cm$^{-3}$, where the flame is essentially
extinguished. We employ a novel numerical representation of the turbulent
flame brush based upon ideas previously developed for modelling laboratory
combustion and explore in some detail the sensitivity of the outcome to the
manner in which burning is initiated. In particular, we simulate 1) a
centrally ignited deflagration, 2) off-center ignition at a single ``point",
and 3) simultaneous off-center ignition at five ``points". We find that the
amount of $^{56}$Ni that is produced and other observable properties depend
sensitively upon how the fuel is ignited. Because of the immediate onset of
buoyant acceleration, the burning region in models ignited off center rises
toward the surface more quickly than in the (commonly assumed) case of
central ignition. With the exception of the model that ignited off-center
at a single point, all models are unbound at the end of the computations and
between 0.59 M$_{\odot}$ (central ignition) and 0.65 M$_{\odot}$   (ignition
at multiple ``points") of matter are processed into nuclear burning
products. These results would guarantee an observable, though weak, Type
Ia supernova. Our results are expected to change for simulations in three
dimensions, especially for the off-center ignitions discussed in this paper,
and late detonations driven by pulsations are not unambigously excluded. We
can, however, state that the chances for a direct transition to a
detonation appear small because, in all our models, the turbulent velocity of
the burning front remains very subsonic.

\end{abstract}


%
\section{Introduction}

The mechanism whereby an accreting carbon-oxygen white dwarf explodes
as a Type Ia supernova continues to be uncertain. While considerable
recent attention has been given to to the so called
``sub-Chandrasekhar mass models'' (e.g.,
\cite{ibetut91,limtor91,kenea93,munren92,woowea94,livarn96}), there remain
compelling reasons to think that a portion, perhaps most Type Ia
supernovae are the explosion of dwarfs that have approached the
Chandrasekhar mass, \mc $\approx$ 1.39 M$_{\sun}$. These reasons include:
1) the recent identification of supersoft x-ray sources (\cite{rapea94})
as the possible progenitors; 2) the 
requirement of conditions that exist only in this class of model for
the nucleosynthesis of the isotopes $^{48}$Ca, $^{54}$Cr, and
$^{50}$Ti (\cite{woowea86,wooeas95}); 3) the
good agreement of the prototypical \mc model, W7 (\cite{nomea84}) with
the observed spectra of typical Type Ia 
supernovae; 4) the good agreement with light curves from this class of
model with a variety of Type Ia supernovae (\cite{hoekho96});
and 5) the difficulty achieving a robust explosion in three
dimensional sub-\mc models (\cite{ben96,garea96}).

Despite this evidence, there remains great uncertainty both as to
whether growth of the CO dwarf to \mc is frequently realized and the
nature of the nuclear explosion once it has begun (\cite{niewoo96}).
The similarity of the progenitor star mass, and therefore of the
initial density structure, suggests a common outcome from this kind of
event. However, there are properties of the progenitor star - the
accretion rate, the original mass of the CO white dwarf before
accretion began, its carbon to oxygen ratio, the metallicity, and
perhaps its rotation rate - that might lead to
diversity. \cite{garwoo95} (GW), in particular, have suggested that
small initial differences might be amplified because the convective
runaway that preceeds the explosion may lead to ignition at multiple
and unpredictable locations. If the global characteristics of the
event, i.e. the ejecta composition and energy, depended strongly on
the ignition conditions some spread of outcomes would follow as a
natural consequence.

Previous multi-dimensional simulations of \mc explosions have
studied the dynamics of {\sl centrally} ignited flame fronts with
perturbed surfaces that first go through the linear Raleigh-Taylor
(RT) instability phase, and later develop rising bubbles of hot, burned
material and turbulence (\cite{muearn86,liv93,kho95,niehill95}). This
assumption of central 
ignition simplifies the calculation and seemed reasonable based
upon existing one dimensional models of the pre-explosive evolution.
Unfortunately, these models all share a common fate - an inadequate amount
of burning occurs to produce a credible explosion. This has led some
to speculate (\cite{kho95,arnliv94a,arnliv94b,niewoo96}) that the
explosion must follow one or more pulsations of the 
white dwarf in which energy is shared between ashes and fuel and a
detonation occurs.

Here, we take a different approach, returning to the simple
(non-pulsing) deflagration models and considering the effects of off
center and multiple point ignition. While our immediate motivation is
the work of GW, there are other lines of reasoning (e.g., \cite{ibe82})
that also suggest off-center ignition. In the GW model ignition points
are set by conditions inside the star when its central temperature
reaches $6 \times 10^8$ K. At this point the time for nuclear burning
and the convective cycle of a burning bubble become comparable. For
certain circumstances, which the star may naturally realize
as it continuously increases its burning rate, rising bubbles of carbon
continue to burn and become more buoyant as they float in an evolution
that is distinctly non-adiabatic.  GW's model suggests that the flame
is born on the surfaces of rising buoyant bubbles with a typical
diameter of about 10 km, at a radial distance between 100 and 200 km
off-center.

To follow these rising burning bubbles, we employ a novel prescription
for flame tracking. Usually the macroscopic burning front is
numerically represented by a thin interface that propagates into the
unburned fuel with a prescribed speed
(\cite{liv93,kho93,kho94,kho95,niehill95}). Here we use a 
``turbulent flame brush'' whose properties are discussed in section
(\ref{num}). Because the prescription is new we calculated both
centrally and non-centrally ignitied flames for comparison.

We find that the amount of $^{56}$Ni that is produced, the overall
energy released, and especially the radial distribution of the burning
products are all quite sensitive to how ignition is
simulated. However, we are unable to conclude unambiguously from the
two-dimensional studies whether the star explodes on the first try or
pulses. Continued studies in three dimesions that calculate the
evolution for a longer time as well as a better understanding of
turbulent flame physics will be necessary to resolve this point.

\section{Numerical representation of the turbulent flame brush}
\label{num}

Owing to the enormous difference between the grid resolution of
multidimensional SN Ia simulations, $\Delta \approx $ 10 km, 
and the thickness of laminar thermonuclear flames, $l_{\rm th}
\approx 10^{-4}$ cm, it is impossible to use the nuclear rate
equations, evaluated at the grid averaged
temperature $\bar 
T_{\Delta}$ and density $\bar \rho_{\Delta}$, to directly compute the
energy generation rate $\dot S$ of a zone that contains fragments of
the turbulent flame brush (``flamelets'') 
(\cite{liv93,kho93,arnliv94b,kho95,niehill95}). In other
words, one has to suitably transform the strongly peaked, unresolved
reaction structure of flamelets immersed in unburned fuel into an average
value for  $\dot S$ of the whole zone. Furthermore, the
propagation speed and thickness of the burning front have to be
modeled in a realistic fashion. Previously, all this has been done by
decoupling the numerical flame front from the grid averaged
thermodynamic quantities and using an artificial scheme to propagate a
thin interface separating fuel and ashes with a prescribed ``turbulent'' flame
speed $u(\Delta)$. The latter can be obtained by invoking isotropic
turbulence, in which case a thick
``turbulent flame brush'' develops that propagates roughly at the
speed of the largest unresolved eddies
$v(\Delta)$  (\cite{niehill95,niewoo96}). 

However, total decoupling from the thermodynamic state of the
numerical cell prevents fluid expansion or compression from directly
affecting the flame. Only by introducing further artificial rules can
local ignition or quenching of the flame by temperature changes be
included in these schemes. An alternative way to treat the problem is
to couple the grid temperature $\bar T_{\Delta}$ to the energy
generation rate via the burning time $\tau_{\rm burn} = \epsilon_{\rm
nuc}/\dot S$. Obviously, if the physical burning times were known in
each zone, the code would be able to decide whether to assume the
presence of highly localized flamelets in the zone, or else to burn
slowly and homogeneously as it does prior to flame ignition. We
therefore state the main assumption of our flame model:
\begin{equation}
\label{assumpt3}
\bar \tau_{{\rm burn}_{\Delta}}(\rho, T, X_{i_{\Delta}}) \approx \tau_{\rm
burn}(\bar \rho_{\Delta}, \bar T_{\Delta},\bar X_{i_{\Delta}})\,\,.
\end{equation}
This assumption is reasonably well satisfied. Immediately after the
flame front enters a grid zone, nuclear energy is deposited into the
zone in the form of internal energy $\bar \epsilon_{{\rm
i}\Delta}$. By virtue of 
the small heat capacity  $c_{\rm p}$ of degenerate matter, reflected
by the numerical equation of state, and the
relation $dT \sim c_{\rm p}^{-1}\,d\epsilon_{\rm i}$, small growth of
$\bar \epsilon_{{\rm i}\Delta}$ already leads to a strong rise of the
zone temperature $\bar T_{\Delta}$. Consequently, $\bar T_{\Delta}$
quickly approaches the final temperature  $T_{\rm b}$ of the burned
mixture, even if only a small part of the zone is burned. The {\it 
numerical} burning time as computed from the grid averaged temperature
(the {\it rhs} of equation \ref{assumpt3}) is thus mostly evaluated at
$\bar T_{\Delta} \approx T_{\rm b}$. This is consistent with the
fact that the {\it physical}, grid averaged burning time (the {\it
lhs} of equation \ref{assumpt3}) is also dominated by burning at
$T_{\rm b}$ due to the strong temperature dependence of $\dot S$. 

In order to distinguish between low temperature, volume filling
burning that is fully resolved by the grid and the high temperature
regime that gives rise to flame formation,  $\tau_{\rm burn}$ must be
related to the characteristic length scale of burning regions. To
this end, we invoke the stationarity assumption $\tau_{\rm burn} =
\tau_{\rm trans}$, where $\tau_{\rm trans}$ stands for the heat transport
time scale (\cite{lanlif91}). Two different transport
mechanisms compete on all
scales: thermal conduction and turbulent transport. Both can be described
by transport coefficients that connect $\tau_{\rm trans}$ (and, hence,
$\tau_{\rm burn}$) to the burning length $l_{\rm burn}$ via 
\begin{equation}
\label{ttrans}
\tau_{\rm cond} = \frac{\rho c_p \,l_{\rm burn}^2}{\sigma}\quad ,
\quad \tau_{\rm tur} = \frac{l_{\rm burn}^2}{\nu_{\rm {tur}}}\,\,.
\end{equation}
The turbulent transport coefficient is the so-called ``eddy
viscosity'' $\nu_{\rm tur} \propto 
v(\Delta) \, \Delta$, where $v(\Delta)$ again denotes the mean magnitude of
turbulent velocity 
fluctuations on the grid scale. One possible technique to compute
$v(\Delta)$ and $\nu_{\rm tur}$ with the help of a subgrid
(SG) model for the unresolved turbulent kinetic energy $q \approx
v(\Delta)^2/2$ was employed  by \cite{niehill95} for SN Ia
calculations; the same method was applied in this work. 
One must be cautious, however, to evaluate the eddy viscosity at the
burning length scale, that is 
\begin{eqnarray}
\label{nulength}
\nu_{\rm {tur}}& =& \nu_{\rm {tur}}(l_{\rm burn}) \nonumber\\
& =&  \nu_{\rm {tur}}(\Delta)\,\left(\frac{l_{\rm
burn}}{\Delta}\right)^{4/3}\,\,, 
\end{eqnarray}
where we have used the Kolmogorov scaling law $v(l) \propto l^{1/3}$.
Turbulent transport is realized on larger scales
while at the same time, thermal conduction dominates small scale
diffusion owing to the different scaling behaviour of both
transport coefficients. Whereas $\nu_{\rm {tur}}$ depends on turbulent
velocity fluctuations that decay on small scales, thermal conduction
by electrons takes over on microscopic scales. The position of this
transition is defined by the burning time (and, therefore, by
temperature). Equations 
(\ref{nulength}) and (\ref{ttrans}) yield the burning lengths at given
$\tau_{\rm burn}$ corresponding to both transport mechanisms:
\begin{equation}
l_{\rm cond} = \sqrt{\frac{\tau_{\rm burn} \sigma}{c_p \rho}} \quad,
\quad l_{\rm tur} = \left(\tau_{\rm burn}\nu_{\rm
{tur}}(\Delta)\right)^{3/2} \, \Delta^{-2}\,\,.
\end{equation}

\begin{figure}
\plotone{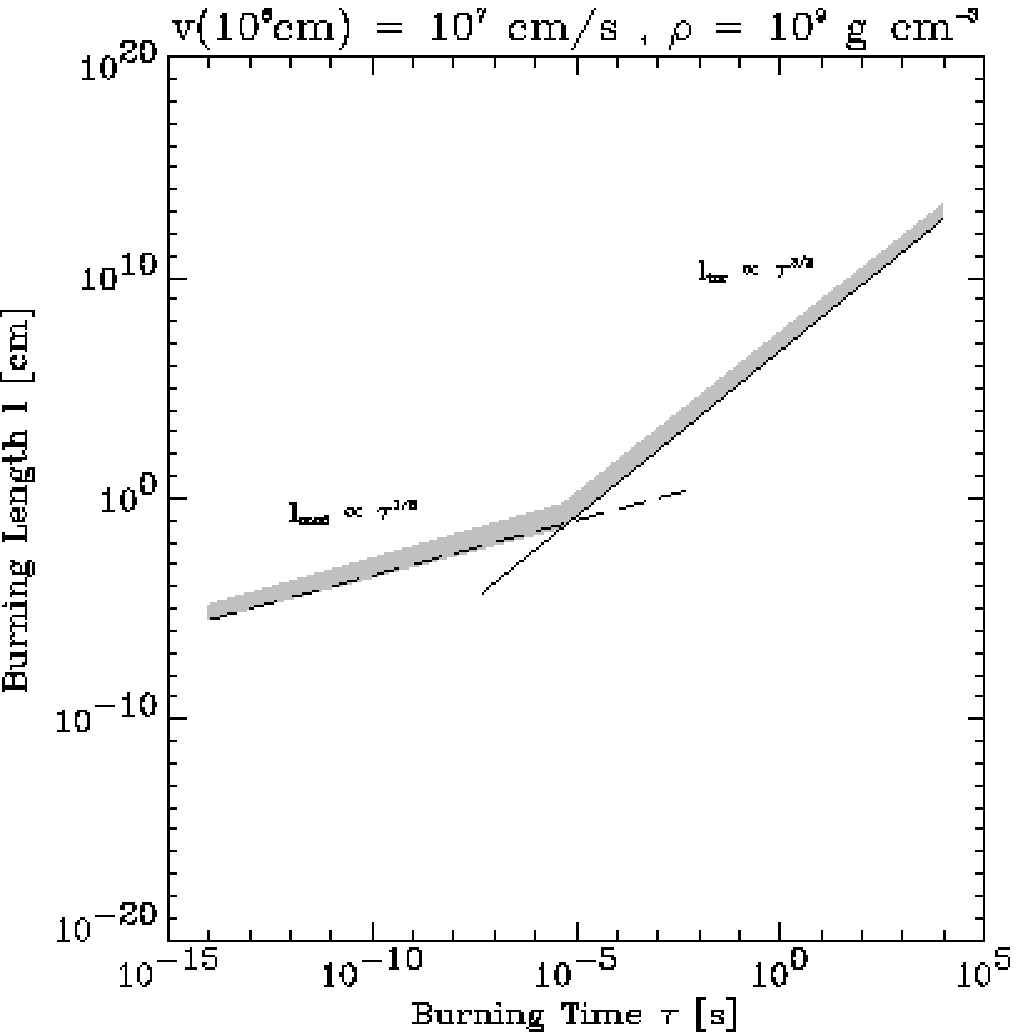}
\caption{\label{lvergl} Scaling of conductive and turbulent length 
scales with the burning time. The shaded region marks the dominant
burning length scale.}
\end{figure}
Fig.~(\ref{lvergl}) demonstrates the behaviour of both quantities at a
typical turbulent intensity and density for thermonuclear flames in white
dwarfs. Since the microscopic burning time at $T_{\rm b} \approx 8
\times 10^9$ K in the flamelet regime is $\tau_{\rm burn} \approx
10^{-12}$ s, the graph shows that microscopic flame dynamics are
governed by thermal conduction.

Phenomenologically, the burning regime that is realized at a
given temperature is the one with the maximum burning length:
\begin{equation}
l_{\rm burn} = {\rm max}(l_{\rm cond},l_{\rm tur})\,\,.
\end{equation}
If $l_{\rm burn} \gtrsim \Delta$, the burning region is fully resolved
and the energy generation rate is given directly by the nuclear
reaction rate. This only occurs prior to ignition of the
zone by the spreading flame brush.
In case of  $l_{\rm burn} < \Delta$, the energy generation rate
can only be computed by means of the effective flame speed
$u(\Delta)$, for reasons explained above. Two further subcases are
possible: if $l_{\rm burn} =  l_{\rm tur}$, the zone is not yet hot
enough to burn in the flamelet regime, as turbulent transport still
dominates at the burning scale. We can approximate the effective
``flame speed'' in this transition regime (corresponding to the
preheating region of laminar flames) by $u(\Delta) \approx l_{\rm
burn}/\tau_{\rm burn}$. As soon as the flame brush fully enters the
zone, the temperature rises to the point where flame transport is governed by
thermal conduction on the scales defined by the burning time,
i.e., $l_{\rm burn} =  l_{\rm cond} \ll \Delta$. The effective {\it
macroscopic} burning
speed then follows from the maximum of all competing transport
velocities at the grid scale:
\begin{equation}
u(\Delta) = {\rm max}(u_{\rm lam}, v(\Delta), v_{\rm rt}(\Delta))\,\,,
\end{equation}
where $u_{\rm lam}$ represents the laminar flame speed, adopted from \cite{timwoo92}. Here, $v_{\rm rt}(\Delta)$ is the asymptotic velocity of
rising buoyant bubbles with the radius $\Delta$:
\begin{equation}
\label{vrt}
v_{\rm rt}(\Delta) = {\cal B}\,\sqrt{g_{\rm eff}\,\Delta }
\end{equation}
(\cite{lay55}), where ${\cal B} = 0.511$ is a constant, $g_{\rm eff} = At\, g$
is the effective gravitational acceleration, and the Atwood number $At
= (\rho_{\rm u} - \rho_{\rm b})/(\rho_{\rm u} + \rho_{\rm b})$ is a
measure for the density contrast of burned (b) and unburned (u)
material. \cite{kho95} confirmed that RT-unstable flames in tubes
with the radius $\Delta$ propagate with a speed given by equation
(\ref{vrt}) with the help of three-dimensional simulations. In our
calculations, however, we used a smaller coefficient (${\cal B} =
0.25$) than the single-bubble value derived by Layzer in order to
account for the presence of many bubbles with smaller radii in one cell.
The turbulent fluctuation velocity $v(\Delta)$ is obtained directly
from the kinetic SG-energy: 
\begin{equation}
v(\Delta) = \sqrt{2  q}\,\,.
\end{equation}
The effective energy generation rate is then calculated from
\begin{equation}
\label{egrnum}
\overline{\left(\rho \dot S\right)}_{\Delta} \equiv \frac{1}{\Delta^3}\int
\limits_{V_{\Delta}}d^3 x\,\rho \dot S \approx \frac{u(\Delta) \bar
\rho_{\Delta} \epsilon_{\rm nuc}}{\Delta}\,\,.
\end{equation}

Assumption (\ref{assumpt3}) enables us to determine the presence of
flame segments in a given zone. Therefore, no artificial prescription
for the thickness of the flame brush is required, since its structure
is now fully determined by the temperature and fuel mass fraction of
the fluid. Moreover, the flame is now allowed to spontaneously
ignite if the temperature increases sufficiently (e.g., by compression)
to yield a burning time that indicates the formation of localized
energy sources. 

In addition to the energy generation rate the model has to provide
proper propagation of
the flame surface. A possible method follows in a very natural way from the
fact that temperature is used as an indicator of flame
presence. The diffusion of internal energy by unresolved turbulent
motions is equivalent to thermal conduction with a modified transport
coefficient which is of the same order of magnitude as the eddy
viscosity. Since the eddy viscosity is known from the SG-model,
turbulent energy diffusion can be parameterized by 
\begin{eqnarray}
\label{etrans}
{\bf h}& =& \rho \, C_{\epsilon_{\rm i}}\nu_{\rm {tur}} \nabla
\epsilon_{\rm i} \nonumber \\
& = &  \rho \, C_{\epsilon_{\rm i}}\nu_{\rm {tur}} c_p \nabla \bar T
\end{eqnarray}
if the specific heat $c_p$ is assumed to be locally constant. The
constant $C_{\epsilon_{\rm i}}$ is of order unity; it has been
calibrated by numerical tests to provide the correct flame speed (see
\cite{nie95} for details). Equation (\ref{etrans}) shows that a
temperature jump spreads over the distance $\Delta$ by
turbulent diffusion during a characteristic time $\tau_{\rm diff}
\approx \Delta^2/\nu_{\rm {tur}} \approx \Delta/ v(\Delta)$. Since the
burning timescale of a turbulent flame on the scale $\Delta$
that moves at a speed of $u(\Delta) \approx v(\Delta)$ is
equally given by $ \Delta/ 
v(\Delta)$ we see that the proposed flame model is self-consistent in
the turbulent burning regime: the boundary of the high temperature
region  moves with the speed of a turbulent flame front.

\section{Simulations of central and off-center ignitions}
\label{models}

The simulations involved an Eulerian PPM-based
code to solve the two-dimensional hydrodynamical
equations. Specifically, we employed the program PROMETHEUS
(\cite{fryea89}). Its 
equation of state included contributions of ideal baryon
and photon gases, and of relativistically degenerate electrons. The
gravitational potential was computed in monopole approximation, while
adding the resulting acceleration in a time symmetric fashion 
provided second order accuracy.

In order to account for turbulent flame propagation, the program was
modified with the SG-model described
in \cite{niehill95}. All calculations were
based on a two-dimensional stationary grid with $256 \times 64$ zones
in spherical ($r, \vartheta$) coordinates. Assuming rotational and
equatorial symmetry, the boundary conditions were chosen to be
reflecting everywhere except at the outer radial edge, where outflow
was allowed. Owing to complications with unnaturally high production of
SG-turbulence at reflecting walls, both the SG-model and thermonuclear
burning were inhibited in close vicinity of the grid boundaries. 
The zones were placed at equidistant radial and angular intervals
with the exception of the 40 outermost ones whose radial separation was defined
to increase at a rate of 10 \% each. A
Cray Y-MP computer at the Rechenzentrum Garching was used for all simulations.

The initial model
represented  a \mc white dwarf with a central density $\rho_{\rm c} =
2.8 \times 10^9$ g cm$^{-3}$ and a central temperature $T_{\rm c} = 7
\times 10^8$ K. In addition to initial temperature and density profiles
a starting value for the 
turbulent kinetic SG-energy  $q_0$ had to be assigned. Earlier experiments
showed that the calculations are rather insensitive to the choice of
$q_0$ since the SG-model adjusts to the strength of local shear on very
short timescales (\cite{niehill95}). A
reasonable value is provided by the typical convective velocities
$v_{\rm conv} \approx 5 \times 10^5$ cm/s on the grid scale $\Delta
\approx 10^6$ cm during the pre-ignition phase; this corresponds to
$q_0 \approx 10^{11}$ erg/g. 

The flame front was initiated by instantly changing
the composition of a certain number of zones to nickel while
increasing their temperature to several billion degrees, equivalent to
raising their internal energy by the binding energy  $\epsilon_{\rm
nuc} \approx 7 \times 10^{17}$ erg/g. Immediately after the simulation
has started pressure equalizes across the interface of burned and
unburned material by emitting weak compression and expansion waves in
opposite directions. Meanwhile, turbulent transport increases the
temperature in unburned matter surrounding the burned regions and
lowers their burning times. After a brief period of burning in the
intermediate regime (no flamelets on small scales, transport dominated
by turbulence on all scales) a conductive flame front forms after
approximately $10^{-3}$ s.

\subsection{Centrally ignited deflagrations}
In the central ignition (CI) model, a spherical region
around the star's center was ignited at $t = 0$ s. Its surface was perturbed
by a periodic pattern with a wavelength of $\pi/4$ and an
amplitude of 25 km; this corresponded to a maximum radial distance of
the initial flame front from the star's center of 150 km. The 
development of the burning front in this and all other models is
visualized by snapshots taken at 0.1, 0.8, and 1.4 s. In
all figures, the velocity field is represented by randomly selected
vector arrows at the respective points, superimposed with contour lines
for the density and shaded areas to indicate the energy generation
rate.

\begin{figure}
\plotone{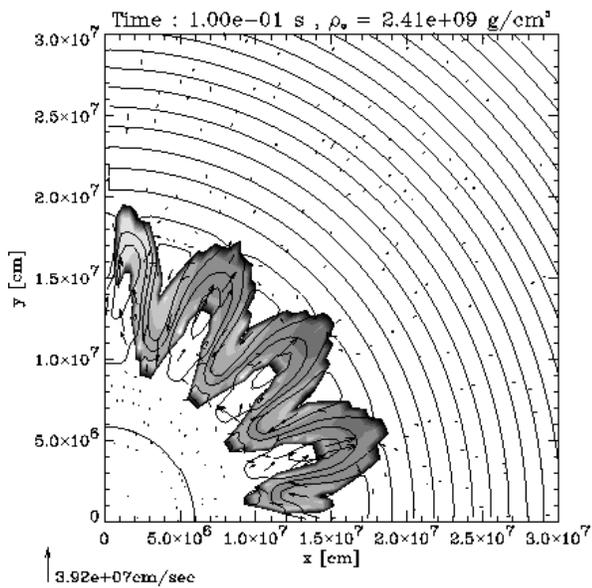}
\caption{\label{cenfl1} Core region in the CI-model shortly after
ignition. Density contours are separated by $\delta \rho = 5 \cdot
10^7$ g cm$^{-3}$, the maximum energy generation rate (shaded regions) is $\dot
S_{\rm max} = 10^{19}$ erg g$^{-1}$ s$^{-1}$. }

\end{figure}
Shortly after ignition, the flame propagates in a direction normal to
its surface and 
forms a pattern that resembles the cellular structure discussed in
\cite{kho95} and \cite{niewoo96}. The cusps pointing toward
the burned matter are clearly visible in fig.~(\ref{cenfl1}). Large
vortices have 
developed, causing the onset of a turbulent cascade that corresponds
to the production of turbulent SG-energy in this model (note that
SG-turbulence is only produced in non-burning
zones). At $t \approx 0.1$ s, the turbulent regime dominates over laminar
propagation almost everywhere within the front.

As the density encountered by the flame brush decreases and effective
gravity rises, buoyancy effects begin to govern the front
evolution. The cellular structure is disrupted by the formation of hot
bubbles floating into the cold material. 

\begin{figure}
\plotone{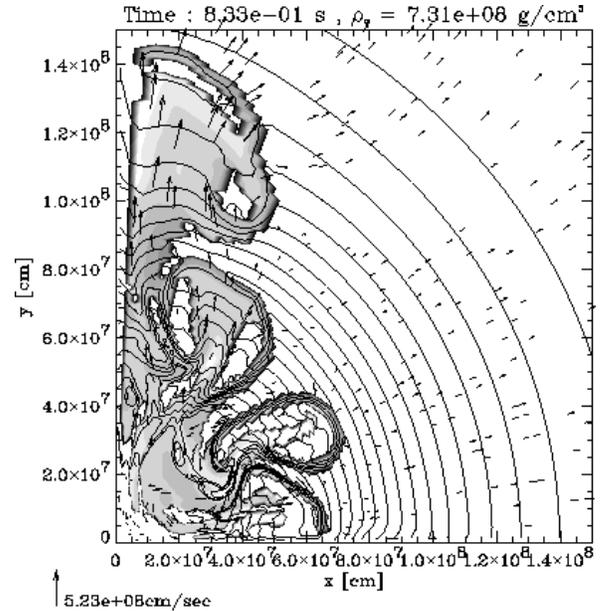}
\caption{\label{cenfl8} $\delta \rho = 3 \cdot
10^7$ g cm$^{-3}$,  $\dot S_{\rm max} = 1.3 \cdot 10^{19}$ erg g$^{-1}$
s$^{-1}$. }
\end{figure}
At $t \approx 0.8$ s, displayed in fig.~(\ref{cenfl8}), the RT-instability of
the burning front has developed to its maximum extent. The 
turbulent flame speed is now $u(\Delta) \approx 2 \times 10^7$ cm/s at
the points of maximum turbulent subgrid energy. Four major
bubbles can be identified, separated by thin streams of unburned
C+O-material. As \cite{kho94} points out, this particular pattern
is an artifact of the two-dimensional simulation, as opposed to an
increased fragmentation of the front in three-dimensional
calculations. Since the material velocity on large scales is
governed by buoyancy and the small scales
behave as if fragmented in three-dimensional turbulence by virtue of
the SG-model, the consequences of this effect may not be
essential. However, as more efficient fragmentation also leads to an
increased production of turbulence it is possible that it indirectly
alters the strength of the explosion. It is still
necessary to confirm this assertion by means of three-dimensional simulations.

After the star has expanded and the density of
the burning region has fallen to some $10^7$ g cm$^{-3}$, the flame is
extinguished at low densities. 
In our code,
this occurs when $l_{\rm burn} \approx \Delta$. At this
stage, incomplete nuclear burning presumably produces intermediate
mass elements that can be observed in SN
Ia spectra. The simple network employed in this work is
unable to predict the final composition of the burning products at
densities below $10^7$ g cm$^{-3}$. Therefore, the exact results of this
stage of our simulations have to be interpreted with some care, since
they can only qualitatively represent the physically realistic
behavior of a dying flame. However, the energetical aspects of the
explosion are widely unaffected by this shortcoming and may therefore
be taken at face value. It is also possible to give an estimate of the
amount of intermediate mass elements produced in our models by keeping
track of the amount of burned matter as a function of burning density (see
section \ref{global}). 
 
\subsection{Off-center ignition at one point}
\label{1B}
GW's arguments suggest that
flame ignition may occur on the surface of buoyant bubbles that have
already risen to radial distances of approximately 200 km . According
to their estimates, the bubbles have diameters of some 10 
km and velocities around 100 km/s at the time of flame formation. In
order to study the effects of off-center ignition we carried out a
number of simulations where the flame was ignited at only one point
located at a radius of 200 km. As only one zone was initially
burned in these one bubble (1B)-models, the radius of the ignited
region roughly equaled $0.5 \Delta 
\approx 5$ km. Some trial calculations involved giving the ignited
zone an initial velocity of 100 km/s, but after very short times the
velocity field relaxed to the same solution as in the case of an
initially resting ignition zone. It should be kept in mind in the
ensuing discussion that the assumption of rotational and equatorial
symmetry enters strongly in this simulation. Despite the fact that the
actual structure of the burning front represents two toroidal buoyant
rings we shall keep the term ``bubble'' in the following description
of the front evolution.

\begin{figure}
\plotone{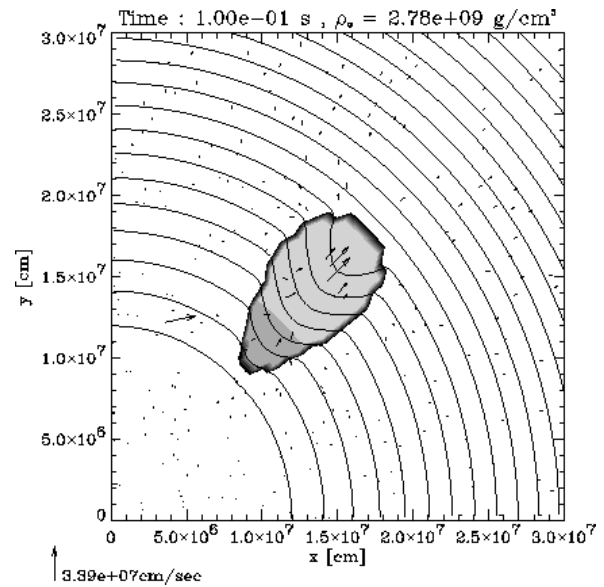}
\caption{\label{onefl1} 1B-model with $\delta \rho = 6 \cdot
10^7$ g cm$^{-3}$ and $\dot S_{\rm max} = 7 \cdot 10^{18}$ erg
g$^{-1}$ s$^{-1}$.}
\end{figure}
The bubble quickly attains a drop-like shape and instantly begins to
float with some $10^7$ cm/s (fig.~\ref{onefl1}). The energy generation
rate is weak compared with the CI-model at the same time, resulting from the
lack of strong shear flows. This can be explained by the larger amount
of potential energy of hot material in the CI-model at the time of
ignition. It is quickly transformed into kinetic
energy as the RT-structures begin to float, giving rise to the 
production of turbulent SG-energy. The floating volume in the
1B-model, on the other hand, is smaller in the beginning and 
therefore unable to provide a large amount of turbulent kinetic
energy to accelerate the burning front.

\begin{figure}
\plotone{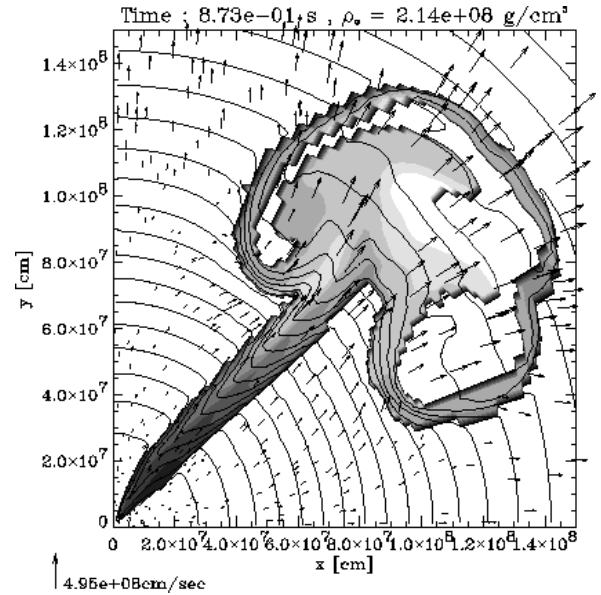}
\caption{\label{onefl8} $\delta \rho =
10^7$ g cm$^{-3}$, $\dot S_{\rm max} = 10^{19}$ erg g$^{-1}$
s$^{-1}$.}
\end{figure}
Having reached a maximum radial extent of approximately 450 km and a rising
velocity of more than $10^8$ cm/s, the bubble begins to spread out in
the angular dimension. fig.~(\ref{onefl8}) shows the burning bubble in
the fully developed 
``mushroom stage'': a nearly spherical head with a diameter of roughly
500 km is attached to a thin pole that connects it to the center of
the expanding star. The most prominent feature of this configuration
is the large volume of burning material at low densities compared to
the part of the front that burns at higher density. It may help to
account for the production of intermediate mass elements necessary to
reproduce SN Ia spectra. A more detailed discussion of this aspect
will follow in section (\ref{global}).


\subsection{Ignition at five points}
\label{5B}
As only little is known about the circumstances of flame ignition it
is also possible that ignition occurs at many points scattered throughout
the white dwarf core region almost simultaneously. The multiple 
ignition model (specifically, ignition at five points (5B)) can be
viewed as a compromise between the 
instantaneous incineration of the core region at $t = 0$ and ignition
at only one point. This way, some desirable properties of both the 
CI-model (strong velocity gradients triggering a powerful turbulent
cascade) and the 1B-model (burning at lower density, increasing the
synthesis of intermediate mass elements) may be combined. 

The model presented here differs from the standard models described
for the CI and 1B-cases with respect to the turbulent diffusion
coefficient  $C_{\epsilon_{\rm i}}$ (section \ref{num}). While
$C_{\epsilon_{\rm i}} = 2$ was used in all other simulations of
turbulence dominated flames this calculation was carried out with
$C_{\epsilon_{\rm i}} = 1$. It therefore models a sligthly more slowly
propagating flame front, while the burning rate remained unchanged. In
order to assess the influence of the propagation and burning rates on
the explosion a second simulation with identical initial conditions
but different calibration parameters ($C_{\epsilon_{\rm i}} = 2$ and
${\cal B} = 0.511$) was carried out; in section (\ref{global}), these
models are referred to as the ``slow'' and ``fast'' 5B-models, respectively.

\begin{figure}
\plotone{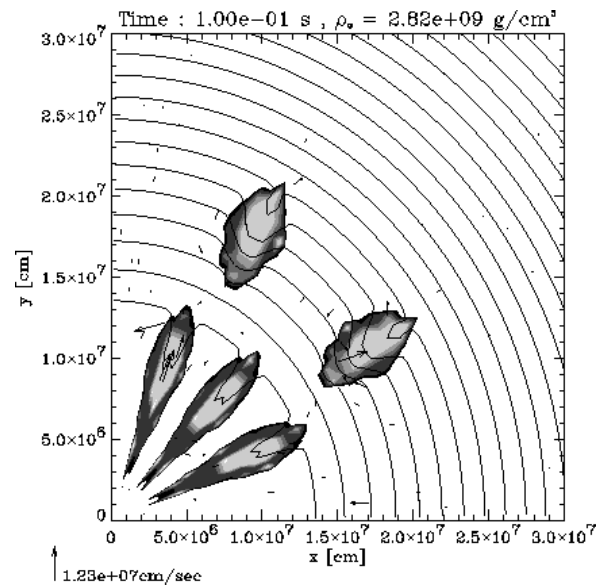}
\caption{\label{5slow1} ``Slow'' 5B-model; $\delta \rho = 6 \cdot
10^7$ g cm$^{-3}$ and $\dot S_{\rm max} = 6 \cdot 10^{18}$ erg
g$^{-1}$ s$^{-1}$.} 
\end{figure}
The initial ignition pattern chosen for the 5B-model can be seen in
fig.~(\ref{5slow1}). Three zones were incinerated at $r \approx 100$
km while two more bubbles were triggered at  $r \approx 200$ km. This
choice follows from the intuitive notion that ignition is more likely
at higher temperatures closer to the center. Initially, all five
bubbles evolve independently, floating in the same way as described for
the 1B-model. After a short time, the bubbles begin to interact. As a
consequence of 
their lower buoyant speed the inner three bubbles merge first, later joined by
by the upper two.

\begin{figure}
\plotone{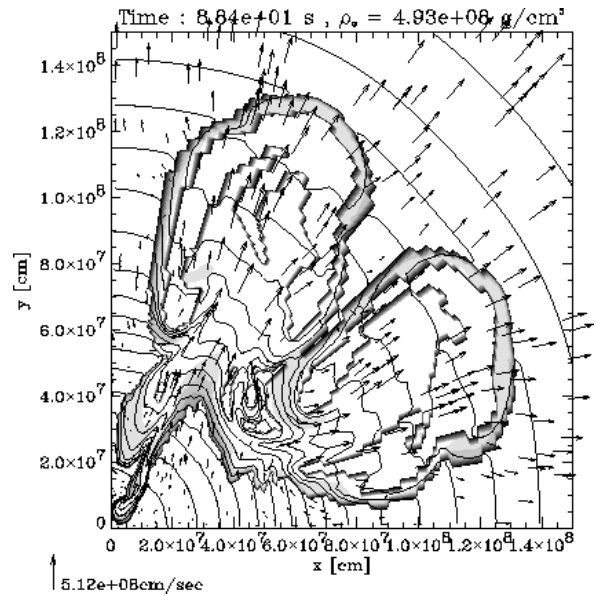}
\caption{\label{5slow8} $\delta \rho = 3 \cdot
10^7$ g cm$^{-3}$, $\dot S_{\rm max} = 10^{19}$ erg g$^{-1}$
s$^{-1}$.}
\end{figure}
It is interesting to compare fig.~(\ref{5slow8}) with
figs. (\ref{cenfl8}) and (\ref{onefl8}). While the maximum material velocity
(shown at the bottom of the plots) of the 5B-model equals that of the
central ignition calculation, its central density is considerably
lower. Furthermore, most of 
the burning material is further off center and therefore burns at a lower
average density, similar to the 1B-model. However, the energy
generation rate in the 5B-model is higher than in the latter, indicating
stronger turbulent motions. 
The apparent similarity of front evolution in both off-center
models and their difference from the CI-simulation gives evidence that
under most natural initial conditions, the bulk volume of burning material can
be expected to float upward. The fact that this is not observed in
central ignition simulations is probably related to the
symmetry conditions that are usually imposed at the grid
boundaries. 

At the time of flame extinction, the high extent of fragmentation
compared with the other models becomes visible.
The flame covers a substantial part of the
picture and still burns strongly in regions of higher density. This
fact explains the higher release of energy in the 5B-explosion
compared with the other models (section \ref{global}). 

\section{Global energetic parameters}
\label{global}

\begin{figure}
\plotone{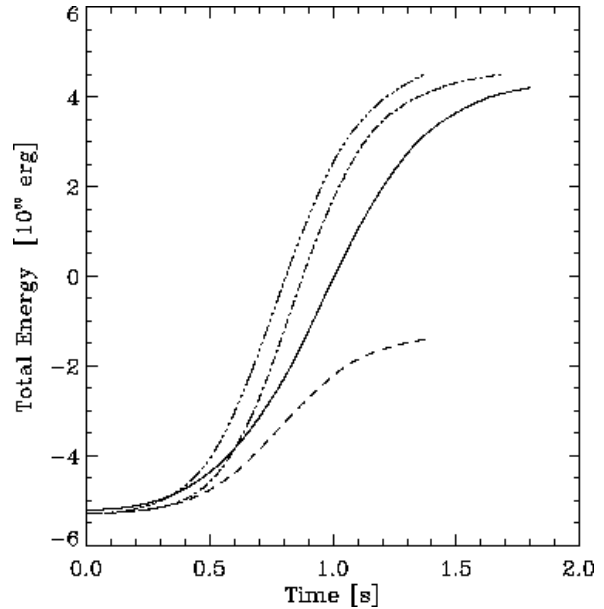}
\caption{\label{etot} Total energy of the different models as a
function of time (--- CI, - - 1B, - $\cdot$ - slow 5B, - $\cdot \cdot
\cdot$ fast 5B).}
\end{figure}

The most important parameter to determine the final fate of the white
dwarf is its total energy. So far, all multidimensional
simulations of the problem have failed to release the required $\gtrsim
0.5$ M$_{\odot}$ in the first deflagration phase needed to
account for SN Ia observations, therefore implying pulsational
detonations if Chandrasekhar mass models were to explain these
events (\cite{kho95,arnliv94a,arnliv94b}). The simulations 
presented in this work provide a possible explosion
mechanism without involving detonations, but the explosions are still
too weak to explain most Type Ia supernovae. This can be seen
in fig.~(\ref{etot}), showing the total
energy as a function of time. All models start at a value of
approximately $-5 \times 10^{50}$ erg, consisting mostly of negative
gravitational energy and positive Fermi energy of the degenerate
electron gas. Note that the CI-model (solid line) has a slightly
higher total energy at $t = 0$ which is related to the larger
initially incinerated mass. 

The only model that remains bound after burning has ceased is the
1B-simulation. It burns less than $0.3$ M$_{\odot}$ in the first
deflagration phase and is thus the
only candidate for a pulsational detonation. According to the
description in section (\ref{1B}), the front fails to ignite a
sufficiently large volume because of its limited volume and the lack of
fragmentation. While the latter aspect may change in a three
dimensional treatment, it must be stressed that the simulation already
assumes rotational and equatorial symmetry. Three-dimensional
simulations of the full star would therefore show a much smaller
volume of the burning bubble and thus predict the generation of
considerably less energy. Hence, it is unlikely that 
one-point ignitions yield enough energy for a healthy explosion on the
first try unless
the burning front spreads significantly faster than modeled. 

One may, of course, speculate about the subsequent evolution of the
1B-model. Some burning material, surrounded by large amounts of
unburned carbon and oxygen, probably remains in the flamelet regime
until the star recontracts. Re-ignition could then occur on a highly
convoluted and 
aspherical surface that results from folding and compression of the
thin stream of burning matter reaching toward the center. Whether this
process is able to form a detonation, or else produces more
buoyant bubbles with turbulent burning surfaces cannot be answered at
this point. In order to obtain information about the re-ignition phase
the multidimensional evolution has to be followed for several seconds
after the initial deflagration, which is computationally unfeasible at
present.

The CI-explosion terminates burning after producing  $0.59$
M$_{\odot}$ of nuclear fuel, which represents the energetical threshold of
an observable, but weak, SN
Ia  event. Despite its higher mass of initially burned material and
the relatively large surface of the burning front it
releases less energy than the 5B-models.

Finally, both fast and slow 5B-simulations appear to represent
possible candidates 
for some Type Ia supernovae. Independent of the variations in burning and
propagation speed mentioned in section (\ref{5B}), between  $0.65$
M$_{\odot}$ and  $0.67$ M$_{\odot}$ matter are burned,
liberating approximately $9 \times 10^{50}$ erg of nuclear energy
(this number is sligthly overestimated due to our assumption of NSE
even at low burning densities, see the next paragraph).

\begin{figure}
\plotone{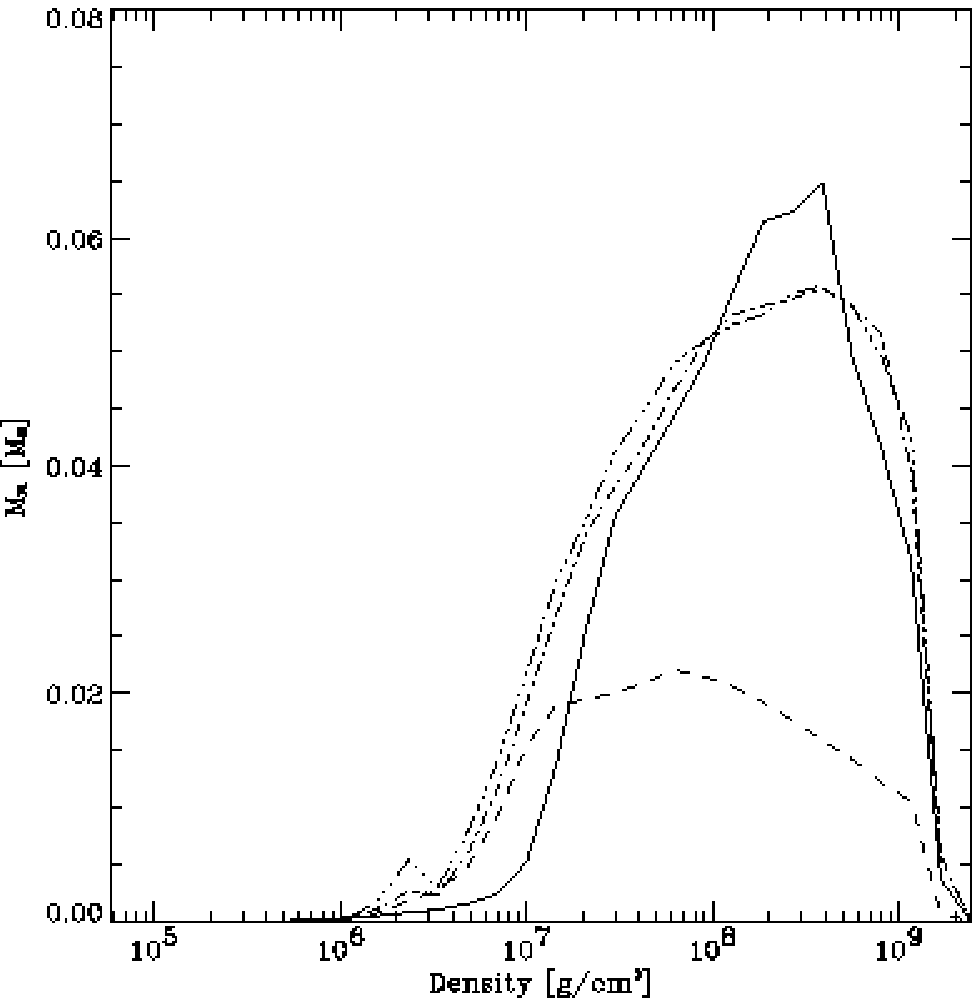}
\caption{\label{brall} Mass of burned matter as a function of the
burning density at $t \approx 1.4$ s (--- CI, - - 1B, - $\cdot$ -
slow 5B, - $\cdot \cdot \cdot$ fast 5B).}
\end{figure}
To conclude the discussion of the performed numerical experiments some
implications for nucleosynthesis in SN Ia events shall be pointed
out. On account of the simplified nuclear network that implies NSE for
the burning products, only indirect statements about the expected final
composition are possible. For this reason, the amount of burned carbon
and oxygen as a function of burning density is plotted in
fig.~(\ref{brall}) for all models. Of particular interest is the
region below $10^7$ g cm$^{-3}$ where intermediate mass elements begin to
be synthesized. The solid curve, representing the
CI-model, shows a sharp peak at approximately $4 \times 10^{8}$
g cm$^{-3}$ followed by a rapid decline toward lower densities. Very
little burning occurs in the intermediate mass region. On the other
hand, all three off-center simulations burn significantly larger
amounts of matter below $10^7$ g cm$^{-3}$. Most surprising is the ratio
of this value to the total amount of burned matter for the 1B-model:
it reaches 20 percent, compared to only four percent for the
CI-model. Additional information about the expected
nucleosynthesis can be gained from the region above $\approx 1.5 \times
10^9$ g cm$^{-3}$. There, the high electron density leads to excess
neutronization compared to solar composition (\cite{woo90}). As
expected, the 1B-model burns the smallest amount of 
C+O-matter at high densities. However, both 5B-models burn even more
than the CI-simulation which might be explained by the rapid initial
expansion of the latter during the first $10^{-2}$ s because of its
higher mass of initially ignited material. 

The total mass of burned material near the end of the simulations
is summarized in table (\ref{tb1}), together with the parameters of
equations (\ref{vrt}) and (\ref{etrans}) used for the different models.

\section{Conclusions}

It can reasonably be assumed
that flame ignition in \mc models for white dwarf explosions occurs
under convective conditions that make 
ignition at the precise center unlikely. Based upon a simple model
for the ignition process (GW), we 
expect that {\it in general}, ignition will occur at  
one or many points within the core. Three models were studied in
two-dimensional simulations: ignition
at one (1B)  and five (5B) points slightly off-center, in addition to
the standard central ignition (CI) scenario. 

The overall appearance of both off-center models is strikingly
different from the CI-case. As the bubble (or its merging
components) rises, it leaves behind a trail of burning material while
the uppermost part grows rapidly. At the
maximum of global energy generation the shape of the burning front
resembles a mushroom whose wide head is supported by a thin pole that
reaches from the center to about a third of the star's radius. Various
processes are responsible for this structure: first, the turbulent
flame speed is much less at the bottom because of the smaller velocity
gradients and shear stress. Therefore, angular spreading at
the pole happens at a smaller rate than in the stirred upper
region. Equally important is the flow pattern around and within the
bubble: displacement of cold material by the rising volume creates
large vortices that diverge at the top and converge at the bottom of
the bubble, thereby focusing the pole and expanding the top.
They also cause the high burning velocities in the upper part by creating
high velocity gradients in the mixing region. Formation of the
macroscopic vortices therefore also determines the radius (and,
consequently, density) at which the bubble begins to spread out. In
the simulations, this happens at approximately one third to one half
of the star's radius. It must be kept in mind that the details of this
development, in
particular the formation of large scale vortices, are probably
different in three-dimensional reality. 

The RT-fingers of the CI-model display similar
characteristics on smaller scales, but never develop the same
elongated mushroom shape. This is at least partly due to the fact that initial
perturbations first
have to overcome the linear RT growth stage before becoming fully
buoyant. By the time the nonlinear evolution
sets in, the dynamics of the largest scales is already strongly
influenced by the expansion of the star. From this point of view,
off-center ignition models simply skip the linear instability regime
and consequently gain important time to rise and become turbulent. The
most important consequence of this behavior is a higher total burning
rate at early times than predicted by standard
deflagration scenarios. We can expect this effect to be even more
pronounced in three dimensions and with a large number of initial
ignition points. If ignition occured almost simultaneously
throughout the central region, but located on bubble surfaces
with small radii compared to that of the ignition region, the bubbles
would reach their maximum burning rate at roughly the same time.
The amount of matter that has been burned by then,
which in turn determines the degree of expansion of the star at the
time rapid burning sets in, will have to
be calculated by three-dimensional simulations with high resolution.

A rough estimate of the average bubble velocity in the limit of many
interacting bubbles in three dimensions can be obtained from experiments
(\cite{rea84,sniand94}) and simulations (\cite{you84,li96}) of RT
mixing fronts in the fully nonlinear regime. It was shown
that the front boundary propagates into the unmixed fluid with the 
velocity 
\be
\label{vfront}
v_{\rm front} = 2\,\alpha g_{\rm eff}\,t\,\,,
\ee
independent of the initial conditions. Here, $t$ is the time since the
beginning of the nonlinear RT stage and $\alpha \approx 0.07$ is a
constant. This behavior was successfully explained by statistical
models pioneered by Sharp and Wheeler (\cite{sha84}) and refined by
\cite{glili88,glisha90} and \cite{zha90}. The Sharp-Wheeler model assumes
that interacting bubbles merge to form bigger ones while small bubbles
are being washed downstream, thus increasing
the average bubble radius $\langle r \rangle$ at the front
boundary. Together with the rise velocity of single bubbles
(\ref{vrt}), this process can account for a constant increase of
$v_{\rm front}$.  It is easily seen that (\ref{vrt}) and
(\ref{vfront}) agree if $\langle r \rangle$ grows linearly with the
distance of the bubbles from the inner front boundary, which is in our
case equal to the size of the burned out region (this is also clear
from dimensional analysis). Furthermore, our bubbles increase
their volume by rapid burning and by gas expansion, providing
additional growth as they rise outward. Hence, equation
(\ref{vfront}) is a reasonable approximation even if bubble merging  is not
the dominant dynamical effect. Using typical numbers 
for white dwarf explosions, $g_{\rm eff} \approx 10^9$ cm s$^{-2}$ and
$t \approx 1 \dots 2$ s, we obtain front velocities of the order of $2
\times 10^8$ \cms. According to one dimensional studies (\cite{woo90})
with similar velocities, this front velocity is sufficient to
produce a healthy explosion.

As a second result of off-center ignition, considerably more matter is 
burned in a lower density environment than in the central ignition 
simulation (table \ref{tb1}). This helps to satisfy two major
constraints on potential SN Ia explosion mechanisms: only little nuclear
burning at high density is permitted to limit the overproduction of
neutron-rich isotops, and a large amount of intermediate mass nuclei
produced at $\rho \lesssim 10^7$ g cm$^{-3}$ is required to explain the
observed spectral lines (\cite{woo90}). Comparing the respective ratios of both
masses and the total burned matter, the 1B-model is clearly the best
candidate in this particular respect. While the 5B-models produce large amounts
of intermediate mass isotops they also burn more high density material
than the CI-model. In the overall comparison of the expected
nucleosynthesis, however, all off-center models are closer to the
expectations from SN Ia observations and solar abundance studies
(\cite{woo90}) than the CI-model.

Looking at the final values for the total energy the
1B-model fails to be a Type Ia supernova
(neglecting the 
possibility of a pulsational detonation). Despite the symmetry
assumptions that imply 
identical energy generation in both hemispheres, too little 
nuclear energy is released to unbind the star. We can therefore
conclude that in order to produce a successful explosion in the first
deflagration phase, flame ignition must occur at a number of points
within a short 
period of time (less than $\approx 10^{-1}$ s when the star begins to
expand), where the term ``points'' 
refers to regions with a diameter equal to or less than 10 km (the
size of our grid zones). Both 5B-simulations and the CI-model become
unbound during their explosions. However, the CI-case would only
represent a weak 
SN Ia since it barely produces the minimum value of approximately 0.6
M$_\odot$ (table \ref{tb1}) of nuclear burning products necessary to
account for light curves and kinetic energy of the ejecta
(e.g., \cite{woo90}). The 5B-calculations, on the other hand, are in
most respects possible candidates for some SN Ia events,
although the final amount of produced $^{56}$Ni is still insufficient to
account for the light curves of the full class of explosions. 
We can thus summarize that deflagration models of \mc white
dwarfs with a small number of burning bubbles are unlikely to produce
powerful SN Ia's unless our simulations 
fail to cover the relevant physics. In addition to the
expected changes in three dimensions described above, examples for 
potentially important, but neglected effects include the additional
production of turbulence by burning itself (\cite{niewoo96}) or from
differential rotation of the star. 

Of course, all results of supernova simulations have to be placed in
the context of their numerical technique. In our case, two 
numerical tools, the SG model for unresolved turbulence and the flame
propagation algorithm, deserve closer inspection as they are both new
and crucial ingredients in our work. The SG model and some of its
numerical properties have been discussed in (\cite{niehill95} and 
references therein). Within its limits (no production of turbulence by
thermal expansion within the flame brush, artificial closure
assumptions) it is capable of representing the unresolved turbulence
in a way that we consider realistic. Dramatic changes in the outcome
of the simulations by changing the SG parameters within reasonable
limits cannot be expected (\cite{niehill95}). The same is true for
the flame algorithm, whose performance has been tested in
\cite{nie95}. However, more effort is necessary to study the problem
of numerically representing turbulent flame fronts in finite
difference schemes like PPM. So far, the turbulent flame brush has
either been treated like a thin, laminar flame surface that moves with
the turbulent burning speed
(\cite{kho95,niehill95}), or has been smeared out over a large region by
turbulent diffusion (this work). Both methods are probably wrong, at
least to some extent. It cannot be excluded that increasing the
numerical resolution and our ability to model turbulent flames will
change our results, although it is unclear in which direction.

The possible transition to a
delayed or, in case of the 1B model, pulsational detonation has not been 
considered in this work but remains, in principle, a viable option for
SN Ia models. However, all of our results show subsonic
effective burning velocities at all times during the turbulent
deflagration phase, making the delayed formation of a detonation very
unlikely. Again, the situation may turn out differently if one
consideres an ensemble of many bubbles that start interacting almost
simultaneously.  

The sensitivity of our results to the initial conditions opens a new
dimension of parameter space that is
worth exploring. One can speculate that some of the variations noticed in
SN Ia observations can be accounted for by this effect. However, if
ignition occurs at sufficiently many points throughout the core region,
the outcome is again expected to be statistically
homogeneous. Detailed studies of the ignition process will be
necessary to answer this question.

\acknowledgements
We would like to express our thanks to S.~Blinnikov and A.~Kerstein
for interesting discussions on flame physics, and to E.~M\"uller for
assistance with the 
numerical experiments. We also thank the referee F.X.~Timmes for helpful
suggestions to improve this manuscript. JCN and WH acknowledge the
hospitality of UCO/Lick Observatories where some of this work was
carried out. JCN was supported by a DAAD HSP II/AUFE fellowship. SEW
was supported by the 
NSF (AST 9115367; AST 94-17161), NASA (NAGW 2525; NAG5-2843), and, in
Munich, by an award from the Humboldt Foundation.

 
\begin{deluxetable}{cccccc}

\tablecaption{Subgrid parameters and burned masses after $t \approx
1.4$ s for all models.  \label{tb1}}

\tablehead{
\colhead{Model} & \colhead{${\cal B}$} & \colhead{$C_{\epsilon_{\rm
i}}$} & \colhead{M$_{\rm b}$[M$_\odot$]}  &
\colhead{M$_{\rm b}$($\rho \gtrsim 1.5 \cdot 10^9$) [M$_\odot$]}
& \colhead{M$_{\rm b}$($\rho \lesssim 10^7$) [M$_\odot$]}} 

\startdata
CI & 0.25 & 2 & 0.59 & $3.5 \cdot 10^{-2}$ & $2.5 \cdot 10^{-2}$ \nl
1B & 0.25 & 2 & 0.27 & $1.1 \cdot 10^{-2}$ & $5.3 \cdot 10^{-2}$ \nl
5B, slow & 0.25 & 1 & 0.65 & $4.5 \cdot 10^{-2}$ & $6.9 \cdot
10^{-2}$ \nl
5B, fast & 0.511 & 2 & 0.67 & $4.8 \cdot 10^{-2}$ & $8.3 \cdot
10^{-2}$ \nl
\enddata
\end{deluxetable}

\newpage


\clearpage

\begin{thebibliography}{}

\bibitem[Arnett \& Livne 1994a]{arnliv94a} Arnett, W. D., \& Livne,
E. 1994a, \apj, 427, 315

\bibitem[Arnett \& Livne 1994b]{arnliv94b} Arnett, W. D., \& Livne,
E. 1994a, \apj, 427, 330

\bibitem[Benz 1996]{ben96} Benz, W. 1996, in Thermonuclear Supernovae,
eds. R. Canal \& P. Ruiz LaPuente, in press.

\bibitem[Fryxell \ea 1989]{fryea89} Fryxell, B.A., M\"uller, E., \&
Arnett, W.D. 1989, MPA Preprint Series, 449

\bibitem[Garcia-Senz, Bravo, \& Woosley 1996]{garea96} Garcia-Senz, D.,
Bravo, E., \& Woosley, S. E. 1996, in Thermonuclear Supernovae, eds.
R. Canal and P. Ruiz-LaPuente, in press

\bibitem[Garcia-Senz \& Woosley 1995]{garwoo95} Garcia-Senz, D., \&
Woosley, S. E. 1995 (GW), \apj, 454, 895

\bibitem[Glimm \& Li 1988]{glili88} Glimm, J., \& Li, X.L. 1988,
Phys. Fluids, 31, 2077

\bibitem[Glimm \& Sharp 1990]{glisha90} Glimm, J., \& Sharp,
D.H. 1990, \prl, 64, 2137

\bibitem[H\"oflich \& Khokhlov 1996]{hoekho96} H\"oflich, P., \& Khokhlov, A.
1996, \apj, in press.

\bibitem[Iben 1982]{ibe82} Iben, I., Jr. 1982, \apj, 253, 248

\bibitem[Iben \& Tutukov 1991]{ibetut91} Iben, I., Jr., \& Tutukov, A. V.
1991, \apj, 370, 615.


\bibitem[Kenyon \ea 1993]{kenea93} Kenyon, S. J., Livio, M., Mikolajewska,
J., \& Tout, C. A. 1993, \apjl, 407, L81


\bibitem[Khokhlov 1991a]{kho91a} Khokhlov, A.M. 1991a, \aap, 245, 114

\bibitem[Khokhlov 1991b]{kho91b} Khokhlov, A.M. 1991b, \aap, 245, L25

\bibitem[Khokhlov 1991c]{kho91c} Khokhlov, A.M. 1991c, \aap, 246, 383

\bibitem[Khokhlov 1993]{kho93} Khokhlov, A.M. 1993, \apjl, 419, L77

\bibitem[Khokhlov 1994]{kho94} Khokhlov, A.M. 1994, \apj, 424, L115

\bibitem[Khokhlov 1995]{kho95} Khokhlov, A.M. 1995, \apj, 449, 695

\bibitem[Landau \& Lifshitz 1991]{lanlif91} Landau, L.D., \& Lifshitz,
E.M. 1991, Lehrbuch der Theoretischen Physik VI: Hydrodynamik (Berlin:
Akademie-Verlag) 

\bibitem[Layzer 1955]{lay55} Layzer, D. 1955, \apj, 122, 1

\bibitem[Li 1996]{li96} Li, X.L. 1996, Phys. Fluids, 8, 336

\bibitem[Limongi \& Tornambe 1991]{limtor91} Limongi, M., \& Tornambe, A.
1991, \apj, 371, 317.

\bibitem[Livne 1993]{liv93} Livne, E. 1993, \apj, 406, L17

\bibitem[Livne \& Arnett 1996]{livarn96} Livne, E., \& Arnett,
W.D. 1996, in Thermonuclear Supernovae, eds. R. Canal \& P. Ruiz
LaPuente, in press 

\bibitem[M\"uller 1994]{mue94} M\"uller, E. 1994, Fundamentals of Gasdynamical
Simulations, MPA Preprint Series, 780

\bibitem[M\"uller \& Arnett 1986]{muearn86} M\"uller, E., \& Arnett,
W.D. 1986, \apj, 307, 619

\bibitem[Munari \& Renzini 1992]{munren92} Munari, U., \& Renzini, A. 1992,
\apjl, 397, L87

\bibitem[Niemeyer 1995]{nie95} Niemeyer, J.C. 1995, Ph.D. Thesis, MPA
Preprint Series, 911 

\bibitem[Niemeyer \& Hillebrandt 1995]{niehill95} Niemeyer, J.C., \&
Hillebrandt, W. 1995, \apj, 452, 769

\bibitem[Niemeyer \& Woosley 1996]{niewoo96} Niemeyer, J.C., \&
Woosley, S.E. 1996, subm. to ApJ 

\bibitem[Nomoto, Thielemann \& Yokoi 1984]{nomea84} Nomoto, K., Thielemann,
F.-K., \& Yokoi, K. 1984, \apj, 286, 644


\bibitem[Rappaport, DiStefano, \& Smith 1994]{rapea94} Rappaport, S., Di
Stefano, R., \& Smith, J. D. 1994, \apj, 426, 692

\bibitem[Read 1984]{rea84} Read, K.I. 1984, Physica D, 12, 45

\bibitem[Sharp 1984]{sha84} Sharp, D.H. 1984, Physica D, 12, 3

\bibitem[Snider \& Andrews 1994]{sniand94} Snider, D.M., \& Andrews,
M.J. 1994, Phys. Fluids, 6, 3324

\bibitem[Timmes \& Woosley 1992]{timwoo92} Timmes, F.X., \& Woosley,
S.E. 1992, \apj, 396, 649 

\bibitem[Woosley 1990]{woo90} Woosley, S.E. 1990, in  Supernovae,
ed. A.G. Petschek (Berlin: Springer-Verlag), p.182

\bibitem[Woosley \& Eastman 1995]{wooeas95} Woosley, S. E., and Eastman, R.
G. 1995, Proceedings of Menorca School of Astrophysics, eds. E.
Bravo, R. Canal, J. Ibanez, and J. Isern, Societat Catalana de Fisica,
p. 105

\bibitem[Woosley \& Weaver 1986]{woowea86} Woosley, S. E., \& Weaver, T. A.
1986, in Radiation Hydrodynamics in Stars and Compact Objects, eds. D.
Mihalas \& K.-H. A. Winkler, (Springer Verlag:Berlin), p. 91

\bibitem[Woosley \& Weaver 1994]{woowea94} Woosley, S. E., \& Weaver, T. A.
1994a, \apj, 423, 371.

\bibitem[Youngs 1984]{you84} Youngs, D.L. 1984, Physica D, 12, 32

\bibitem[Zhang 1990]{zha90} Zhang, Q. 1990, Phys. Lett. A, 151, 18

\end{thebibliography}
\end{document}